\documentclass[twocolumn]{revtex4}
\usepackage{graphicx}
\usepackage{amssymb}
\usepackage{amsmath,amsfonts,epsfig}
\usepackage[english]{babel}
\usepackage{graphicx}

\newcommand{\kT}{k_{\rm{B}}\mathrm{T}}
\newcommand{\rvec}{\textbf{r}}
\newcommand{\zunity}{\hat{z}}
\newcommand{\lb}{\lambda_{\rm{B}}}

\newcommand{\Figs}{Figs.}
\newcommand{\fig}{Fig.}
\newcommand{\Fig}{Fig.}
\newcommand{\Eqs}{Eqs.}
\newcommand{\eqs}{Eqs.}
\newcommand{\Eq}{Eq.}
\newcommand{\eq}{Eq.}
\newcommand{\y}{\mathsf{y}}

\begin{document}
\title{Charge regulation and ionic screening of patchy surfaces}
\author{N. Boon and R. van Roij}
 \address{Institute for Theoretical Physics, Utrecht University, Leuvenlaan 4, 3584 CE Utrecht, The Netherlands}

\begin{abstract}
The properties of surfaces with charge-regulated patches are studied using non-linear Poisson-Boltzmann theory. Using a mode expansion to solve the non-linear problem efficiently, we reveal the charging behaviour of Debye-length sized patches. We find that patches charge up to higher charge densities if their size is relatively small and if the patches are well separated. The numerical results are used to construct a basic analytical model which predicts the average surface charge density on surfaces with patchy chargeable groups.
\end{abstract}
\maketitle
\section{Introduction}

Most surfaces that are immersed in an aqueous solution obtain a net charge due to ion adsorption or dissociative processes at the surface. The resulting electrostatic force between such surfaces is essential for understanding the stability, osmotic pressure and flocculation behaviour of colloidal suspensions. In the standard (linear) screening picture, like-charged surfaces repel each other at distances of the order of the Debye screening length,\cite{DLVO1, DLVO2, Barrat} due to overlapping clouds of screening ions in the vicinity of the two surfaces. This electrostatic repulsion, combined with short-range attractive Van der Waals  forces if there is a dielectric contrast between the colloidal particles and the solvent, is a basic result of Derjaguin-Landau-Verwey-Overbeek (DLVO) theory which dates back to the 1940's.\cite{DLVO1, DLVO2, Barrat} More recent studies, based on nonlinear Poisson-Boltzmann theory, also find strictly repulsive electrostatic forces between pairs of like-charges surfaces.\cite{Neu, Grunberg, Klein, Tamashiro, Tellez, Trizac2} Nevertheless, there are also experimental reports of attraction between like-charge colloids at ranges much longer than those of Van der Waals forces. Ion-ion correlations, which are ignored in the mean-field type Poisson-Boltzmann theory, might be an explanation for these observed attractions in the case of multi-valent ions.\cite{Linse, Netz, Nguyen} However, evidence for electrostatic attractions has also been reported for suspensions with only monovalent ions,\cite{Ise, Kepler, Carbajal-Tinoco, Larsen} causing heated debates in the literature on the break-down of the classic DLVO theory due to many-body effects, the vicinity of glass walls, hydrodynamic forces, etc. Interestingly, it has also been suggested that {\em charge inhomogeneities} can be responsible for these attractions,\cite{Vreeker, Chen, Perkin, Popa} where the heterogeneity of the surface charge may be due to an incidentally present or purposely designed underlying chemical structure, or by clustering of adsorbed surfactants. \\

Apart from the ill-understood electrostatic attractions in some systems, another good reason for considering heterogeneously charged surfaces in more detail stems from the fascinating recent advances in the chemical synthesis of a large class of novel patchy nanoparticles, featuring not only corners, edges, and facets due to their finite size, but also spots or stripes.\cite{Glotzer} Understanding the large-scale self-assembly properties of these newly available nanoparticles is an important ongoing scientific quest that requires effective particle-particle interactions as an input. For this reason a better understanding is needed of the relations between the chemical heterogeneity of patchy particles, the resulting surface charge density, and the nature and geometry of the ionic screening cloud that ultimately dictates the effective electrostatic interactions. In this article we explore some of these relations within nonlinear Poisson-Boltzmann (PB) theory in the relatively simple geometry of chargeable stripes on a planar surface in contact with a bulk electrolyte.

In PB theory a key role is being played by the boundary conditions (BC's), in particular those on the surface between the suspending electrolyte and the suspended colloidal nanoparticles. The most common BC is to predescribe the surface charge density of the colloidal particle, thereby imposing a fixed discontinuity of the displacement field at both sides of the surface while the surface potential itself increases upon the approach of another like-charged surface. This type of constant-charge BC was shown to be realistic for e.g. strongly acidic homogeneous surfaces that ionise completely in a polar solvent such as water.\cite{Behrens2} An alternative is to predescribe the electrostatic potential on the surface, such that the particles can adjust their charge density if another surface approaches. This constant-potential BC has turned out to be realistic for surfaces not too far from a point of zero charge.\cite{BehrensReg} It is not clear, however, which of these BC's is realistic for surfaces with a heterogeneous chemical composition. For instance, it is not trivial how the charge of a highly charged patch on an otherwise weakly charged or neutral surface is distributed, and how the induced electrostatic potential propagates to affect the charging of nearby surface groups of other patches. The problem requires us to treat the interplay between the electrostatic potential and the surface charge density at the more microscopic level of a chemical equilibrium of attaching and detaching ions as modeled by charge-regulation BC's.\cite{Popa2,BehrensReg,Pericet} Such an approach has been applied to fit force measurements between two homogeneously charged plates,\cite{BehrensReg, Pericet,Popa2} or to compute forces between periodically modulated charged plates within linearised PB theory.\cite{Miklavic1} Here, however, we combine charge regulation BC's with nonlinear PB theory for a single spatially heterogenous surface. Using a newly developed numerical scheme we will reveal how surfaces with finite-sized discrete patches\cite{Miklavic1,Richmond,Khachatourian} tend to charge up and get screened, where the full non-linear PB theory allows us to deal with highly charged patches next to oppositely charged or neutral areas with strongly varying ion concentrations perpendicular and parallel to the surface.

\section{Theory}

\begin{figure}[]
\centering
\includegraphics[width = 8cm]{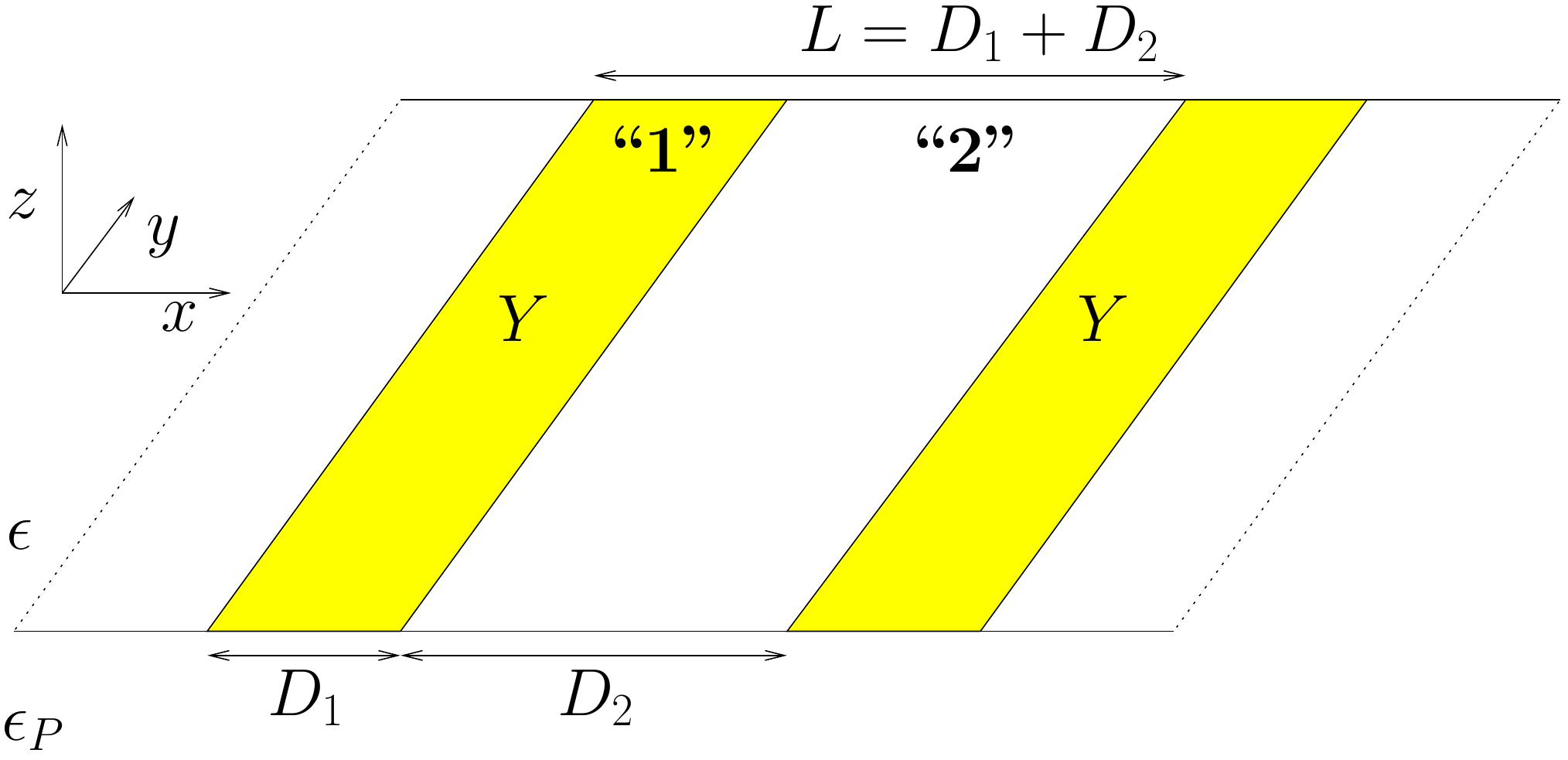}
\caption{Pictorial representation of the model. The striped patches are shaded to distinguish these from the rest of the plate and will be referred to as region ``1''. The area in between the patches will be referred to as region ``2''.} \label{fig:MODEL1}
\end{figure}

We consider a solid medium in the half space $z<0$, characterized by a dielectric constant $\epsilon_P$.  The surface of this medium at $z=0$ is considered to be chemically heterogeneous, e.g. with a stripe pattern such as depicted in \fig~\ref{fig:MODEL1}. The half space $z>0$ is a bulk solvent with dielectric constant $\epsilon$ and volume $V$ at temperature $T$. This solvent is assumed to be in thermal and diffusive equilibrium with a reservoir at $z\rightarrow \infty$ that contains point-like monovalent cations and anions, both at a concentration $\rho_s$. Since treating the ions in a mean-field fashion is in most of the cases an accurate approach in the case of monovalent ions in water or the more polar (higher $\epsilon$) oils,\cite{Punkkinen, Valeriani2} we describe the distributions of ions for $z>0$ by the Boltzmann distributions $\rho_{\pm}(\rvec) = \rho_s \exp(\mp \Phi(\rvec))$. Here $\kT \Phi(\rvec)/e$ is the electrostatic potential at $\rvec = (x,y,z)$, with $e$ the elementary charge and $\kT$ the Boltzmann constant. We assume that $\Phi(x,y,z\rightarrow\infty)=0$ in the reservoir, and that the densities of both ion species are zero for $z<0$ due to hard-core repulsions. The Poisson equation relates the charge density to the Laplacian of the potential, $\nabla^2 \Phi(\rvec) = -4\pi\lb (\rho_+(\rvec) - \rho_-(\rvec))$, where we introduce the Bjerrum length $\lb = e^2/\epsilon \kT$. What follows is the Poisson-Boltzmann (PB) equation,
\begin{eqnarray}
\!\!\!\!\nabla^2 \Phi(\rvec) &=&\left\{\begin{array}{ll}\kappa^2\sinh\Phi({\bf r})&z\geq0;\\ 0&z<0,\end{array}\right.\label{PB}
\end{eqnarray}
where the Debye screening length is given by $\kappa^{-1} = (8\pi\lb \rho_s)^{-1/2}$. At the fluid-solid interface, the presence of a surface charge density $e\sigma(x,y)$ and a stepwise change in dielectric medium gives rise to a boundary condition
\begin{eqnarray}
\nabla\Phi(x,y,0^+)\cdot\zunity&=&-4\pi \lambda_B\sigma(x,y) \nonumber\\
 &+& \frac{\epsilon_{P}}{\epsilon} \nabla\Phi(x,y,0^-)\cdot\zunity. \label{BC1}
\end{eqnarray}
Other boundary conditions ensure that the potential is continuous at $z=0$, vanishes at infinity, and require that the electric field in the medium vanishes far away from the interface:
\begin{eqnarray}
\Phi(x,y,0^+) &=& \Phi(x,y,0^-);\label{BCcont}\\
\lim_{z\rightarrow \infty} \Phi(x,y,z) &=& 0;\label{BCposinf}\\
\lim_{z\rightarrow -\infty} \Phi{'}(x,y,z) &=& 0. \label{BCmininf}
\end{eqnarray}
Here, and below, a prime denotes a partial derivative w.r.t. the $z$-coordinate.
\subsection{Charge regulation}
The majority of models which apply PB theory assume either a fixed charge density or a fixed potential at surfaces. In reality the charging is often regulated; neither the charge density nor the surface potential is fixed, since both depend on the local density of ions in the surrounding liquid. There is a subtle interplay as the local surface charge density depends on the local surface potential and vice versa.

We consider $4$ major charge generation mechanisms, involving cationic/anionic adsorption/desorption,
\begin{subequations}
\begin{eqnarray}
\mathrm{S} + \mathrm{H}^{+} \rightleftarrows \mathrm{SH}^{+}&\mathrm{~with~}&K^A_+=\frac{[\mathrm{S}][\mathrm{H}^+]}{[\mathrm{SH}^+]}
,\label{reaca}\\
\mathrm{SB} \rightleftarrows \mathrm{S}^{+} + \mathrm{B}^{-}&\mathrm{~with~}&K^D_+=\frac{[\mathrm{S}^+][\mathrm{B}^-]}{[\mathrm{SB}]}
,\label{reacb}\\
\mathrm{S} + \mathrm{B}^{-} \rightleftarrows \mathrm{SB}^{-}&\mathrm{~with~}&K^A_-=\frac{[\mathrm{S}][\mathrm{B}^-]}{[\mathrm{SB}^-]}
,\label{reacc}\\
\mathrm{SH} \rightleftarrows \mathrm{S}^{-} + \mathrm{H}^{+}&\mathrm{~with~}&K^D_-=\frac{[\mathrm{S}^-][\mathrm{H}^+]}{[\mathrm{SH}]}
,\label{reacd}
\end{eqnarray}
\end{subequations}
where S denotes a surface group, $\mathrm{H}^+$ represent a cation, and $\mathrm{B}^-$ an anion, with concentrations $[\mathrm{S}]$, $[\mathrm{H}^+]$, and $[\mathrm{B}^-]$ in the vicinity of the surface. \Eqs~(\ref{reaca}) and (\ref{reacc}) describe associative charging in which an ion from the liquid is adsorbed at the surface. \Eqs~(\ref{reacb}) and (\ref{reacd}) describe dissociative charging, where ions are released from the surface into the liquid. These chemical reactions are characterized by reaction constants $K_\pm^{A/D}$. We assume the charging to occur at discrete sites at the surface, and define a surface site density $m^A_{\pm}(x,y)$ and $m^D_{\pm}(x,y)$ for associatively and dissociatively chargeable sites respectively. Since each surface site is either charged or neutral, we have
\begin{subequations}
\begin{eqnarray}
~[\mathrm{S}]+[\mathrm{SH}^+]&=& m^A_+(x,y),\label{densa}\\
~[\mathrm{S}^+]+[\mathrm{SB}]&=& m^D_+(x,y),\label{densb}\\
~[\mathrm{S}]+[\mathrm{SB}^-]&=& m^A_-(x,y),\label{densc}\\
~[\mathrm{S}^-]+[\mathrm{SH}]&=& m^D_-(x,y).\label{densd}
\end{eqnarray}
\end{subequations}
Using the Boltzmann relations $[\mathrm{H}^+]=\rho_s \exp[-\Phi(x,y,0)]$ and $[\mathrm{B}^-]=\rho_s \exp[+\Phi(x,y,0)]$ for the ion densities at the surface $z=0$, we can express the associative and dissociative surface charge densities $\sigma_{\pm}^{A} = [\mathrm{SH}^+]$, $[\mathrm{SB}^-]$ and $\sigma_{\pm}^{D} = [\mathrm{S}^\pm]$,  respectively, as
\begin{subequations}
\begin{eqnarray}
\sigma^A_\pm(x,y) &=&  \frac{m_\pm^A(x,y)}{1+K^A_\pm/\rho_s\exp(\pm\Phi(x,y,0))} \label{sigmaA};\\
\sigma^D_\pm(x,y) &=&  \frac{m_\pm^D(x,y)}{1+\rho_s/K^A_\pm\exp(\pm\Phi(x,y,0))}\label{sigmaD},
\end{eqnarray}
\end{subequations}
for both types of charging. Here, $\Phi(x,y,0)$ is the surface potential, which for heterogeneous surfaces is a function of the lateral coordinates $x$ and $y$. Note that \emph{all} surface chemistry is encoded in $m^{A/D}_{\pm}(x,y)$ and equilibrium constants $K_\pm^{A/D}$, which we consider as input parameters in this work.\\

For simplicity we only consider stripe-like surface inhomogeneities in this work, such that the charge and the surface potential only depend on $x$ and not on $y$, see \fig~\ref{fig:MODEL1}. Translational invariance in the $y$-direction follows directly by assuming $m^{A/D}_{\pm}(x,y)=m^{A/D}_{\pm}(x)$. Note that the potential  $\Phi(x,y,z)=\Phi(x,z)$ does depend on a lateral coordinate and a normal one.\\

The common  case that only a small fraction of the sites charges ($\sigma \ll m$) necessarily corresponds to the case that the unity in the denominators of \eqs~(\ref{sigmaA}) and~(\ref{sigmaD}) is negligible compared to the other term. In other words, for all cases that the surface is only charged to a fraction of its maximum it is safe to omit the unity from \eqs~(\ref{sigmaA}) and~(\ref{sigmaD}). For most associative reactions this approximation is fine, as was already argued by Grahame in 1947.\cite{Grahame} For dissociating surfaces one should be more cautious, especially at low ionic strengths $\rho_s$.\cite{Healy} In the remainder of this study we employ this assumption throughout, as it reduces the number of independent parameters vastly. This is immediately seen by introducing the dimensionless total surface charge density
\begin{equation}
\y(x) = 4\pi\lb \kappa^{-1} \left(\sigma^A_+(x) + \sigma^D_+(x) -\sigma^A_-(x) - \sigma^D_-(x)\right), \label{totsurfacecharge}
\end{equation}
which reduces \eq~(\ref{sigmaA}) and~(\ref{sigmaD}) within the small-charge-fraction limit to
\begin{equation}
\y(x) = Y^+(x) \exp(-\Phi(x,0)) - Y^-(x) \exp(+\Phi(x,0)), \label{totcharging}
\end{equation}
where
\begin{equation}
Y^\pm(x) = \kappa \left(\frac{m_\pm^A(x)}{2 K_\pm^A} +  \frac{m_\pm^D(x) K_\pm^D}{2 \rho_s^2}\right). \label{chargeabilitydef}
\end{equation}
Note that \eq~(\ref{chargeabilitydef}) defines a dimensionless quantity which is not necessarily of the order of unity. Depending on the material properties and ion concentrations, both $Y^+(x)$ and $Y^-(x)$ can easily vary over many decades. From now on we refer to $Y^+(x)$ and $Y^-(x)$ as the positive and negative \emph{chargeability}, respectively. \Eqs~(\ref{BC1}),~(\ref{totsurfacecharge}) and~(\ref{totcharging}) give rise to the boundary condition for our model,
\begin{eqnarray}
\Phi{'}(x,0^+)&=&\frac{\epsilon_{P}}{\epsilon} \Phi{'}(x,0^-) + Y^-(x) \exp(+\Phi(x,0))  \nonumber\\
&-& Y^+(x) \exp(-\Phi(x,0)). \label{BCdiss}
\end{eqnarray}
The latter equation together with \eqs~(\ref{PB}) and~(\ref{BCcont})-(\ref{BCmininf}) forms a closed set of equations to solve the electrostatic potential above and within the solid medium.

\subsection {The homogeneous limit}
To the best of our knowledge, there are no analytic solutions to the nonlinear Poisson-Boltzmann (PB) equation for inhomogeneously charged surfaces. However, it can be solved analytically in the case of a single homogeneously charged plate, see for example Ref.~\cite{Hunter}. Considering a homogeneous surface charge density $\sigma(x,y)=\sigma_H$, the potential can be written as
\begin{eqnarray}
\!\!\!\!\Phi_H(z) &=&\left\{\begin{array}{ll}2 \displaystyle \ln \frac{1+\gamma \exp[-\kappa z]}{1-\gamma \exp[-\kappa z]}&z\geq0;\\\Phi_H(0)&z<0,\end{array}\right. \label{phihom}
\end{eqnarray}
where $\gamma = (\sqrt{4+\y_H^2}-2)/\y_H$, and $\y_H=4\pi\lb\kappa^{-1}\sigma_H$ is the dimensionless surface charge density as defined in \eq~(\ref{totsurfacecharge}), now for a homogeneously charged plate. Note that the solid medium is free of electric fields in this case, and because of this, \eq~(\ref{phihom}) is independent of the dielectric constant of the plate, $\epsilon_P$. By using \eq~(\ref{totcharging}) for the mechanism to account for associative and dissociative charging we find an expression for the surface potential of homogeneously chargeable plates. In general, for a (positive) homogeneous chargeability $Y^+(x)= Y_H$ (and $Y^-(x)\equiv0$) we obtain from \eqs~(\ref{totcharging}) and~(\ref{phihom}) that the surface potential is given by
\begin{widetext}
\vspace*{-0.5 cm}
\begin{eqnarray}
\Phi_{H}(0) &=& 2 \ln \left\{\begin{array}{ll}
\frac{2}{\sqrt{3}} \cos \left(\frac{1}{3} \arctan
\left(\sqrt{\frac{4}{27 Y_H^2}-1}\right)\right)&~\rm{if}~0\leq Y_H<\frac{2}{3\sqrt{3}};\\
\frac{\frac{2}{3}^{1/3}}{\left(9Y_H+\sqrt{3}\sqrt{27 Y_H^2 -4}\right)^{1/3}} + \frac{\left(9Y_H+\sqrt{3}\sqrt{27 Y_H^2 -4}\right)^{1/3}}{18^{1/3}}&~\rm{if}~Y_H\geq\frac{2}{3\sqrt{3}}.\end{array}\right. \label{phihom2}
\end{eqnarray}
\end{widetext}
The regime $Y_H<1$ is associated with the regime where linear Poisson-Boltzmann theory holds, as we can easily check that surface potentials do not exceed unity here. By taking the inverse of \eq~(\ref{phihom}) we obtain a relation which is well-known from Gouy-Chapman theory,
\begin{equation}
\y_{H} = 2 \sinh \frac{\Phi_{H}(0)}{2}, \label{yhom}
\end{equation}
and this can be used in combination with \eq~(\ref{phihom2}) to find the explicit relation $\y_H(Y_H)$ between charge and chargeability of a homogeneous plate. The limiting cases are
\begin{equation}
\y_H\approx \begin{cases}Y_H&~\rm{if}~Y_H \ll 1;\\
Y_H^{1/3}&~\rm{if}~Y_H \gg 1. \label{limitingcases}
\end{cases}
\end{equation}
Physically this means that the charging of a weakly chargeable surface is relatively efficient, whilst a highly chargeable surface only gains charge with the cube root of the density of chargeable sites. Both regimes of \eq~(\ref{limitingcases}) can easily be distinguished in a double-logarithmic plot; the solid line in \fig~\ref{fig:AVCHARGE} consists of two straight lines with slopes $1$ and $1/3$, respectively, with a cross-over between both regimes at $Y_H\approx 1$.

\section{Mode expansion}
We develop a numerical scheme to solve the PB equation in the case of a striped patchiness on the surface, described by $\sigma(x,y)=\sigma(x)$, such that the charge distribution on the surface only changes in the $\hat{x}$ direction. We assume inhomogeneities to be periodic with period $L$, such that $\sigma(x+L) = \sigma(x)$, and we can thus write
\begin{equation}
\sigma(x) = \sum_{k=-M}^{M} \sigma_k \eta_k(x),
\end{equation}
where $\eta_k(x)\equiv \frac{1}{\sqrt{\kappa L}}\exp(\frac{i 2\pi k x}{L})$ with $k\in\mathbb{Z}$ are conveniently normalised (Fourier-) modes and $\sigma_k$ are the corresponding amplitudes in the expansion, given by $\sigma_k = \kappa \int_{0}^{L} \sigma(x) \eta_k(-x)~\mathrm{d}x$. Note that this charge density is not known beforehand, as it depends on the surface potential through \eq~(\ref{totcharging}). The number $M$ signifies a high-wavenumber cut-off, that we will empirically choose to be large enough to describe the essential large-wavelength physics. We will develop a method to calculate the electrostatic potential $\Phi(x,z)$ at and above the plate, $z\geq 0$. Since this is a function which will be subject to the $L$-periodic symmetry as a function of $x$, such that $\Phi(x,z) = \Phi(x+L,z)$ for all $x,z$, it can be written as
\begin{equation}
\Phi(x,z) = \sum_{k\in\mathbb{Z}} \phi_k(z) \eta_k(x), \label{modeexp}
\end{equation}
where $\phi_k(z)$ is the mode amplitude of the Fourier component $\eta_k(x)$.
One easily checks that functions of the form $\phi_k(z) \eta_k(x) = c_k \exp (\frac{|k|Lz}{2\pi}) \eta_k(x)$, with $c_k$ a constant, solve the PB equation for $z<0$, and satisfy boundary condition~(\ref{BCmininf}). This solution can be put into \eq~(\ref{BC1}), to yield boundary conditions in terms of the mode amplitudes given by
\begin{eqnarray}
\epsilon \phi_k{'}(0^+) &=& -4\pi \epsilon \lambda_B \sigma_k  + \frac{2\pi |k|}{L} \epsilon_P \phi_k(0); \label{fourierbc1}\\
\lim_{z \rightarrow \infty} \phi_k(z) &=& 0. \label{BCmode2}
\end{eqnarray}
The task is now to find PB-like differential equations for every mode $\phi_k(z) \eta_k(x)$ in the regime $z>0$ by inserting the mode expansion~(\ref{modeexp}) into the PB equation~(\ref{PB}). The left hand side can be treated easily, yielding
\begin{equation}
\nabla^2 \Phi(\rvec) = \sum_{k\in\mathbb{Z}} \left(\phi_k{''}(z) - \left(\frac{2 \pi k}{L}\right)^2 \phi_k(z)\right) \eta_k(x), \label{linmodepb}
\end{equation}
where a prime denotes a derivative with respect to $z$. The $\sinh\Phi({\bf r})$ on the right side of the PB equation is a nonlinear function, which gives rise to the couplings between mode amplitudes $\phi_k(z)$ for all $k\in\mathbb{Z}$.\cite{Boon1} This complicates the calculation of the solution of these modes; in general it will not be possible to find solutions for every mode separately.
\subsection{Linear PB}
\begin{figure}[]
\centering
\includegraphics[width = 8cm]{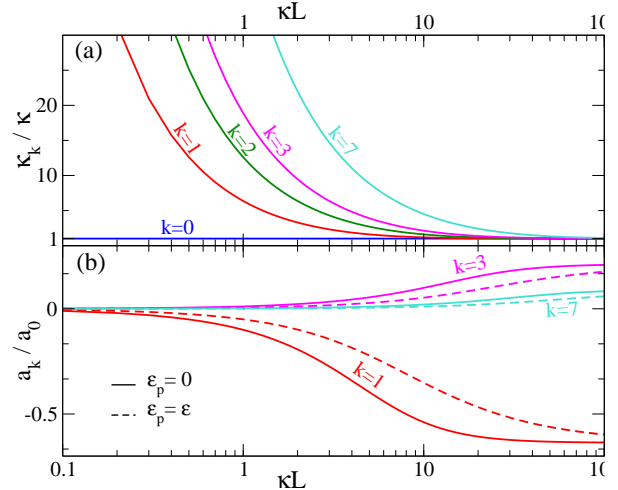}
\caption{The mode-dependent screening parameter $\kappa_k$ (a) and mode amplitudes $a_k$ (b) for several $k$ as a function of the periodicity parameter in the system, in (b) for charged ($y=1$) and uncharged ($y=0$) stripes of equal width. The solid lines in (b) correspond to the case $\epsilon_P = 0$, whilst the dashed lines show data for $\epsilon_P=\epsilon$. } \label{fig:SCREENINGPARAM}
\end{figure}

Only in the case that we describe a weakly charged system, such that the potential is small, and thus $\sinh \Phi({\bf r})\approx\Phi({\bf r})$, the PB equation becomes linear, $\nabla^2 \Phi(\rvec) \approx \kappa^2 \Phi(\rvec)$. In this case the modes decouple, and using \eq~(\ref{linmodepb}) the solution for each mode is found to be
\begin{equation}
\phi_k(z) = a_k \exp(-\kappa_k z) + b_k \exp(\kappa_k z),~~(z\geq0) \label{linearform}
\end{equation}
where $a_k$ and $b_k$ are integration constants and $\kappa_k = \sqrt{\kappa^2 + \left(\frac{2\pi k}{L}\right)^2}$ is the mode-dependent screening parameter. Note that in the present case of a single plate the coefficient $b_k$ vanish because of \eq~(\ref{BCmode2}), and we find by applying \eq~(\ref{fourierbc1}) the amplitudes
\begin{equation}
a_k = \frac{4\pi\epsilon\lb\sigma_k L}{\epsilon \kappa_k L + 2\pi|k| \epsilon_P}.
\end{equation}
\Fig~\ref{fig:SCREENINGPARAM}a shows the relation between the screening parameters $\kappa_k$ and the (dimensionless) periodicity of the system $\kappa L$. Besides the independence of $\kappa_0$ on $\kappa L$, it shows that $\kappa_k \gg \kappa_0$ for small $\kappa L$.
This means that wave-like inhomogeneities in the electrostatic surface potential always vanish within a few wavelengths normal to the surface. Therefore, inhomogeneities with short wavelengths are screened over shorter distances than a Debye length.
From \fig~\ref{fig:SCREENINGPARAM}a and from the definition of $\kappa_k$ it can be concluded that all inhomogeneities in the potential must have essentially disappeared at distances of the order of $L$ from the surface as $\kappa_k \gg1/L$ for all $|k|\neq0$.\\

\Fig ~\ref{fig:SCREENINGPARAM}b shows some of the corresponding mode coefficients $a_k$, for $k=1,3,$ and $7$, all for $\epsilon_P=0$ (non-penetrating fields) and $\epsilon_p=\epsilon$ (index-matched solvent and plate). In the figure we divided by $a_0$ to show the inhomogeneous amplitudes relative to the homogeneous background. All coefficients can be calculated by using the $\sigma_k$ following from a surface with charged (y=1) and uncharged(y=0) stripes of equal width. This is a fixed charge density and therefore we do not account for association/dissociation reactions at the surface here. The coefficients $a_k$, and therefore the inhomogeneities in the potential, are relatively small but may become large for systems with a larger periodicity ($\kappa L \gg 1$). This is analogous to the fact that at small periodicities (with respect to $\kappa^{-1}$) the potential is not able to laterally adapt to the oscillations in the surface charge and becomes more homogeneous, whilst at very high $\kappa L$ it is able to take the form of a step function. The choice of the ratio of the dielectric constants between the plate and the liquid is of importance. We see from \Fig ~\ref{fig:SCREENINGPARAM}b that if we choose an index-matched plate and solvent, corresponding to the situation that the fields are able to penetrate into the solid medium, the coefficients $a_k$ become significantly lower (in absolute value) than if this ratio is chosen close to zero, where we do not find any fields in the medium. In the $\epsilon_P=0$ case, inhomogeneities in the charge distribution will have a larger effect on the inhomogeneity of the associated electrostatic potential, as the polarizability of the solid medium is not able to compensate any inhomogeneities in the electrostatic potential from within the plate.

\subsection{Nonlinear PB}
In the general nonlinear case, the mode amplitudes will not be exponential functions of the distance like in \eq~(\ref{linearform}). Instead we now use $\sinh{\Phi} =  \left(\exp[\Phi] - \exp[-\Phi]\right)/2$ to write the mode expansion of the electrostatic potential as
\begin{equation}
\sinh\Phi({\bf r}) = \frac{1}{2} \prod_{j=-M}^M E_j^+(x,z) - \frac{1}{2} \prod_{j=-M}^M E_j^-(x,z),\label{Eprod}
\end{equation}
where $M$ is the high-wavenumber cut-off and $E_j^\pm(x,z) = \exp \left[\pm \phi_j(z) \eta_j(x)\right]$ contains the nonlinear dependence of the PB equation on $\phi_j(z)$. Using the Taylor expansion of the exponential function, and applying that $\eta_j(x)^n = (\kappa L)^{(1-n)/2} \eta_{n\cdot j}(x)$, we write $E_j^\pm(x,z)$ as a mode sum, where high-frequency modes $\eta_{n\cdot j}$ for which $|n\cdot j|>M$ have been neglected. Note that because of this truncation we only need to expand $E_j^\pm(x,z)$ up to order $M/j$ (and for $j=0$ we choose to stop at order $M$.) As an illustrative example, $E_1^\pm(x,z)$ can be expanded to a mode sum as
\begin{eqnarray}
E_1^\pm(x,z) &=& 1 \pm \phi_1(z) \eta_1(x)\nonumber \\
&+& \frac{\phi_1(z)^2 \eta_2(x)}{2\sqrt{\kappa L}} \pm \frac{\phi_1(z)^3 \eta_3(x)}{6\kappa L} +\dots
\end{eqnarray}
Using the mode-expansion representation of $E_j^{+}(x,z)$, it is now a straightforward task to calculate both the products of all $E_j^{+}(x,z)$ and all $E_j^{-}(x,z)$, as appearing in \eq~(\ref{Eprod}), and rewrite these products as \emph{new} mode sums with mode amplitudes $V_k(z)$,
\begin{eqnarray}
\sinh\Phi({\bf r}) &=& \frac{1}{2} \sum_{k=-M}^{M}\left(  V_k^{+}(z) -  V_k^{-}(z) \right)\eta_k(x) \nonumber \\
&+&\mathcal{O}(\eta_{M+1}(x)). \label{modesumV}
\end{eqnarray}
Note that $V_k^{\pm}(z) = V_k^{\pm} (\phi_0(z),\phi_1(z),\dots,\phi_M(z))$, such that combining \eqs~(\ref{modesumV}) and~(\ref{linmodepb}) yields
\begin{eqnarray}
&&\phi{''}_k(z) - \left(\frac{2 \pi k}{L}\right)^2 \phi_k(z)\nonumber\\
 &=& \frac{\kappa^2}{2}V_k^+(\{\phi_j(z)\}) -\frac{\kappa^2}{2}V_k^-(\{\phi_j(z)\}), \label{nonlinmodepb}
 \end{eqnarray}
for each $k$, $|k|\leq M$. We solve \eq~(\ref{nonlinmodepb}) for all $k$ iteratively as follows. For a given mode-amplitude $\phi_k(z)$ we project out the dependence on all other modes $\phi_j(z)$, $j\neq k$, by expanding $V_k^{\pm}(z) = \sum_{i=0}^{M} U^{\pm}_{k,i}(z) \phi_k^i(z)$ and write \eq~(\ref{nonlinmodepb}) as
\begin{eqnarray}
\phi_k{''}(z) &-& \left(\frac{2\pi k}{L}\right)^2\phi_k(z) =\nonumber \frac{\kappa^2}{2} \sum_{i=0}^{M}U^+_{k,i}(z) (\phi_k(z))^i \\
 & & - \frac{\kappa^2}{2}\sum_{i=0}^{M} U^-_{k,i}(z) (\phi_k(z))^i,~~z>0. \label{modePB}
\end{eqnarray}
As an example we calculate the explicit expressions for the monopole ($k=0$),
\begin{equation}
U^\pm_{0,i} = \frac{(\pm \eta_0)^i }{ i!} \cdot \left(\sqrt{\kappa L} -\frac{\phi_{-1} \phi_1}{\sqrt{\kappa L}} + \frac{\phi_{-1}^2\phi_{1}^2}{4(\kappa L)^{3/2}}  + \dots\right),
\end{equation}
where we omitted higher modes $|k|>1$ and left out the $z$ dependence for brief notation. If we insert this expressions into \eq~(\ref{modePB}) again, we find
\begin{equation}
\phi_0{''}\eta_0 = \kappa^2  \left(1 -\frac{\phi_{-1} \phi_1}{\kappa L} + \frac{\phi_{-1}^2\phi_{1}^2}{4(\kappa L)^2}  + \dots\right) \sinh \left[\phi_0 \eta_0\right],
\end{equation}
which reduces to the planar PB-equation in the case that $\phi_{\pm 1}$ vanish.
\Eq~(\ref{modePB}) together with the BC's~(\ref{fourierbc1}) and~(\ref{BCmode2}) can be used to find the numerical solution to every mode, given the approximations to the solutions for other modes.  In order to obtain the solution of the full problem, we therefore apply an iterative scheme in which we go through multiple cycles of solving each PB mode equation consecutively, until we find the converged solution. This is done by solving for every $\phi_k(z)$ the mode equations~(\ref{modePB}) on a 1-dimensional grid. As a starting point we use a vanishing solution for all modes. It is possible to reduce the computational effort significantly by considering systems for which $\Phi(x,z) = \Phi(-x,z)$, such that all $\phi_k(z)$ take values in the real space $\mathbb{R}$ and $\phi_k(z) = \phi_{-k}(z)$. The number of iterations required for convergence depends on the degree of nonlinearity in the system and is typically in the order of $10$ per mode.
\section{Results}
\subsection{Charging of stripes}
The theory and numerical method we introduced can be used to calculate the charging of any periodic configuration of chargeable parallel stripes on a planar surface. Here, we focus on a plate coated with stripes with periodicity $L=D_1 + D_2$, where $D_1$ and $D_2$ are the widths of the striped regions with dimensionless chargeabilities $Y_1^\pm$ and $Y_2^\pm$ respectively, such that
\begin{equation}
Y^\pm(x) = \begin{cases}Y_1^\pm &\mathrm{if}~x \in \mathrm{region~1};\\
Y_2^\pm &\mathrm{if}~x \in \mathrm{region~2}. 
\end{cases}
\end{equation}
Regions ``1'' and ``2'' are also shown in  \fig~\ref{fig:MODEL1}. We focus mostly on $Y_1^+ \equiv Y$, $Y_1^-=Y_2^\pm=0$, and $\epsilon_{P}/ \epsilon \ll 1$, which corresponds to uncharged areas (regions ``2'') separating stripes with positive chargeability $Y$ on a low-epsilon plate in contact with a high-epsilon liquid such as water. For this particular calculation it suffices to choose $M=64$ modes, a $z$-grid of $N=2000$ points extending to $z=5\kappa^{-1}$.\\

\begin{figure}[]
\centering
\includegraphics[width = 8cm]{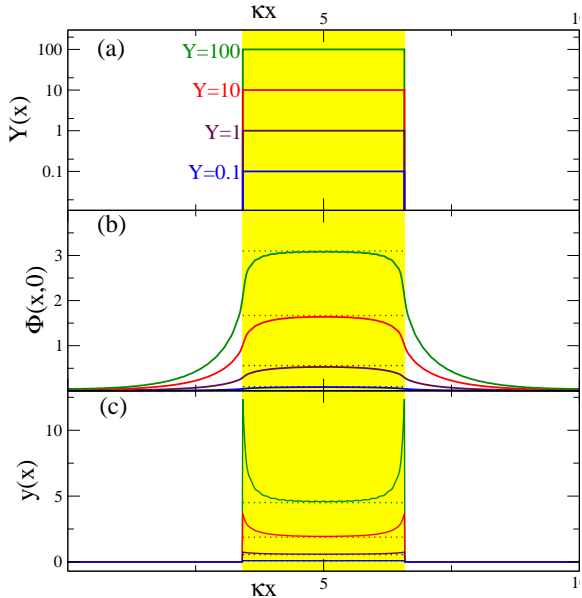}
\caption{Positional (lateral) dependence of (a) the chargeability, (b) the surface potential and (c) the dimensionless surface charge density of a striped patchy surface with periodicity $L=10\kappa^{-1}$ and patch stripe width $D_1=3.1 \kappa^{-1}$, for several stripe chargeabilities. The shaded area shows the position of the stripe on the plate, and the horizontal dotted lines in (b) and (c) show values we would obtain for infinitely wide stripes.} \label{fig:POTPLOT1}
\end{figure}

\begin{figure}[]
\centering
\includegraphics[width = 8cm]{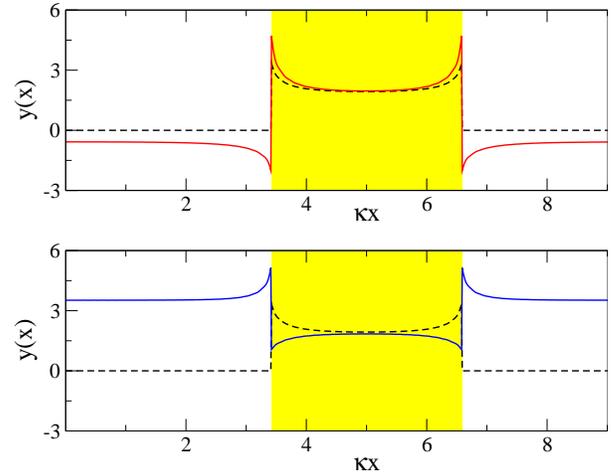}
\caption{Positional (lateral) dependence of the dimensionless surface charge density of a striped patchy surface with periodicity $L=10\kappa^{-1}$ and stripe width $D_1=3.1 \kappa^{-1}$. We indicated the position of the stripe by the shaded areas. Both (a) and (b) show a dashed line which shows data corresponding to $Y=10$ as in \fig~\ref{fig:POTPLOT1}(c), whilst the solid lines show the effect of setting the chargeability of the surrounding plate to (a) $Y_2^-=1$ in combination with $Y_2^+=0$ and (b) $Y_2^+=50$ (in combination with $Y_2^-=0$.)} \label{fig:CHARGEDENSITY}
\end{figure}

\begin{figure}[]
\centering
\includegraphics[width = 8cm]{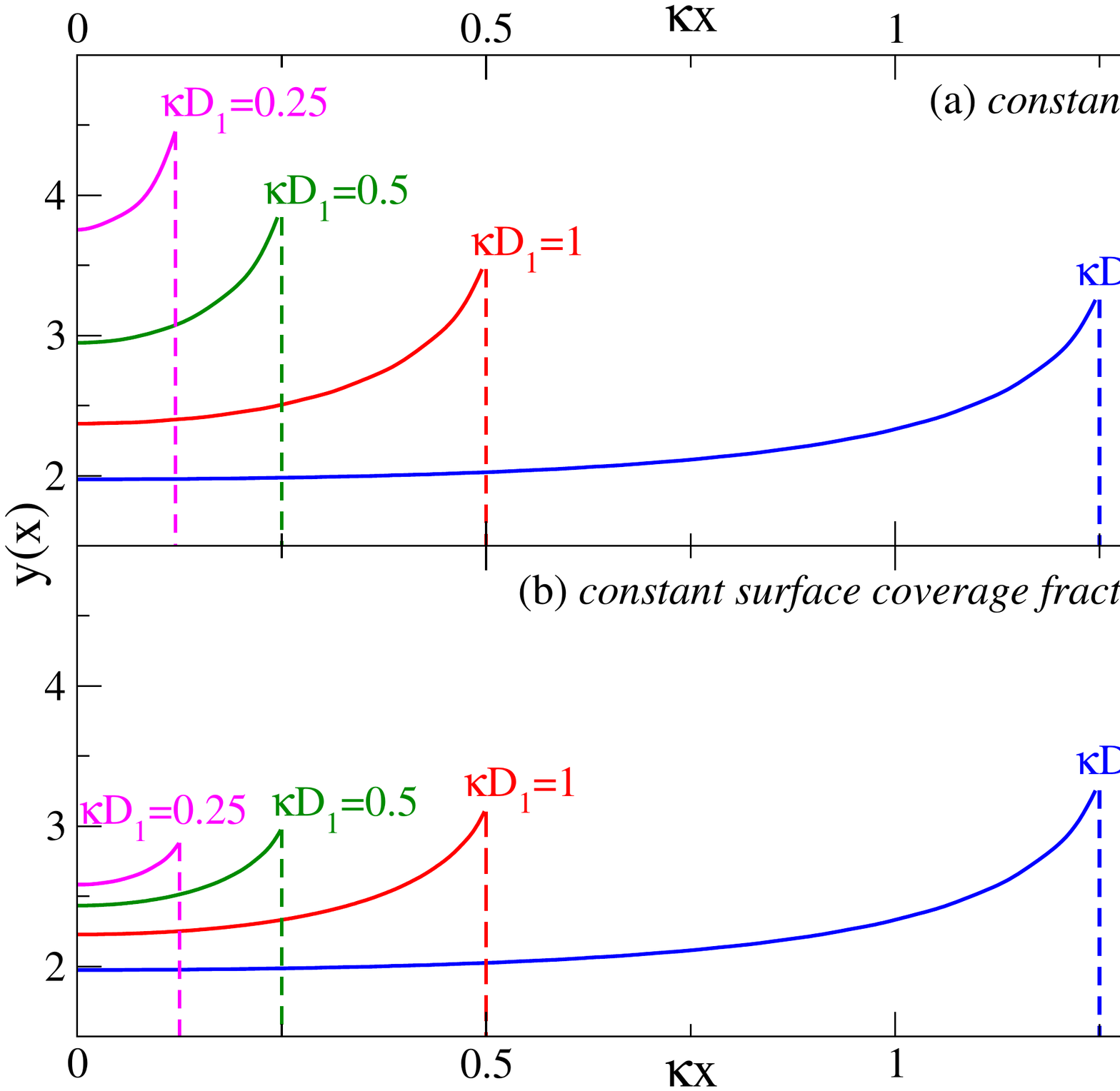}
\caption{Surface charge density profiles $\y(x)$ of a striped patchy surface with $Y=10$ for various stripe widths $D_1$ and spatial periodicities $L$, here as a function of the distance to the center of the charged stripe. In view of the symmetry we only plot half a period. The data in (a) is calculated using the fixed value of $\kappa L=10$, whilst in (b) we use $L = 2D_1$. The vertical dashed lines indicate the edges of the stripes, beyond which $\y(x)=0$.} \label{fig:CHARGEDENSITY2}
\end{figure}

For a periodicity of $10$ Debye lengths, $\kappa L=10$, a charged stripe of width $D_1=3.1 \kappa^{-1}$ (such that $D_2=6.9 \kappa^{-1}$), and several chargeabilities $Y \in \{0.1, 1, 10, 100\}$, Fig.~\ref{fig:POTPLOT1}(a) shows the $x$-dependence of the chargeability, which is a step function. This is the starting point of our calculation, and we calculate the electrostatic potential $\Phi(x,z)$ via the iterative scheme that was described above. \Fig~\ref{fig:POTPLOT1}(b) shows the resulting surface potential $\Phi(x,z=0)$ for the four values of $Y$ of \fig~\ref{fig:POTPLOT1}(a). For each $Y$, the thin horizontal dotted lines show the surface potential in the case that the stripe would have infinite width ($D_1\gg \kappa^{-1}$) by using the analytical solution to the planar-PB \eqs~(\ref{phihom2})-(\ref{yhom}), i.e. the surface potential of a homogeneously charged plate. In the present system the homogeneous surface potential limits the actual surface potential from above. Since the stripes are relatively broad ($D_1>\kappa^{-1})$,  the calculated surface potential approaches this limit, which we will call the homogeneous limit from now on, at the center of the stripe. Typical length scales over which the potential varies laterally are clearly of the order $\kappa^{-1}$, as expected. \Fig~\ref{fig:POTPLOT1}(c) shows the dimensionless surface charge $\y(x)$, which is related to the surface potential via \eq~(\ref{totcharging}). We see that the charge density at the center of the stripe is well described by that of a homogeneously charged plate. By contrast, a relatively high charge builds up close to the edges of the stripe. For $Y\gg1$, in the nonlinear screening regime, the charge density at these edges largely exceeds the values we find at the center. The reason is the nearby neutral surface, which results in a surface potential which is lower at the edge of the stripe than at its center. The charging of the surface groups, which is normally limited by the induced rise of the electrostatic potential according to \eq~(\ref{totcharging}), can therefore be stronger close to the edges. As a result, the average charge density of a stripe can be much higher than what one would expect for homogeneously charged plates with the same chemical properties. The reason that we only observe this effect (deeply) in the \emph{nonlinear} regime is because the surface potentials must be significant($\Phi(x)\geq 1$), such that the Boltzmann factors which govern the surface charge distribution, see \eq~(\ref{totcharging}), deviate strongly from unity.\\

We checked that the results we obtain do not depend on the finite grid size and the number of included modes characterized by $N$ and $M$, respectively; even the curves obtained with $M=16$ and $N=1000$ are indistinguishable from all those in \fig~\ref{fig:POTPLOT1}. It should be noticed that in the case we choose $M$ too small, instead of giving inaccurate results, the iterative scheme does often not converge anymore, such that no solution is found at all.\\

The two panels in \fig~\ref{fig:CHARGEDENSITY} each show the charge density $\y(x)$ for $Y=10$ as was calculated in \fig~\ref{fig:POTPLOT1}(c) for $\kappa D_1=3.1$ and $\kappa L=10$ by a dashed line, while the full curve denotes $\y(x)$ in the case of a modified parameter set. It illustrates cases where the surrounding surface is chargeable as well. \Fig~\ref{fig:CHARGEDENSITY}(a) shows a situation where the surrounding stripe is able to charge up slightly negatively (with chargeability $Y_2^-=1$). The presence of strong peaks of (opposite) excess charge at the edges of the stripes demonstrates that this presence of a chargeable surface with opposite sign of charge enhances the charging of the adjoining area.
If the surrounding stripe is chosen to be more positively chargeable $(Y_2^+ = 50)$ than the original one, as in \Fig~\ref{fig:CHARGEDENSITY}(b), the stripe and the surrounding surface change roles. At the interface between the two regions the potential now is higher than at the center of the stripe, having a value somewhere in between the homogeneous limits of both stripes. This causes the charge density to peak just outside the original stripe whilst inside it shows sharp minima at the edges. \\

The distribution of charge at the chargeable stripe depends on the width and the spatial periodicity of the stripes. This is depicted in \fig~\ref{fig:CHARGEDENSITY2}, where the charge density $\y(x)$ for $Y=10$ is plotted like in \fig~\ref{fig:POTPLOT1}(c), now for various $D_1$ and $L$. \Figs~\ref{fig:CHARGEDENSITY2}(a) and (b) show the effect of changing the width of the stripes but deal differently with the size of region in between the stripes. In \fig~\ref{fig:CHARGEDENSITY2}(a) the stripe periodicity is kept fixed at $L=10\kappa^{-1}$, such that the stripe-fraction of the surface increases with increasing stripe width $D_1$. In \fig~\ref{fig:CHARGEDENSITY2}(b) the stripe periodicity is set to $L=2D_1$, for several $D_1$, such that the surface coverage fraction of the stripes remains fixed at $50\%$. The numerical results in \fig~\ref{fig:CHARGEDENSITY2}(a) clearly show that smaller stripes at a fixed stripe periodicity $L$ gain a higher charge density. From \eq~(\ref{totcharging}) it follows that the maximum charge density will occur for infinitesimally thin stripes, since the surface potential at the stripe will vanish in that case. In this limit, the corresponding charge density is $\y(x) = Y$, for all $x$ at the stripe. The results for a constant surface coverage in \fig~\ref{fig:CHARGEDENSITY2}(b)  fraction also show that thin stripes gain the highest average charge density, although the maximal charge density at the edges decreases a little. The increase is less pronounced than in \fig~\ref{fig:CHARGEDENSITY2}(a) since now the width of the charge-neutral region in between the stripes scales with the width of the stripes. Therefore, thin stripes in \fig~\ref{fig:CHARGEDENSITY2}(b) are relatively close to their neighboring stripes, and will hinder each other in gaining charge. One can show that that the maximum charge density will again be found for infinitesimally thin stripes, for which the system reduces to an essentially homogeneously chargeable plate with chargeability $Y_H = Y D_1/L$.\\

\begin{figure}[]
\centering
\includegraphics[width = 8cm]{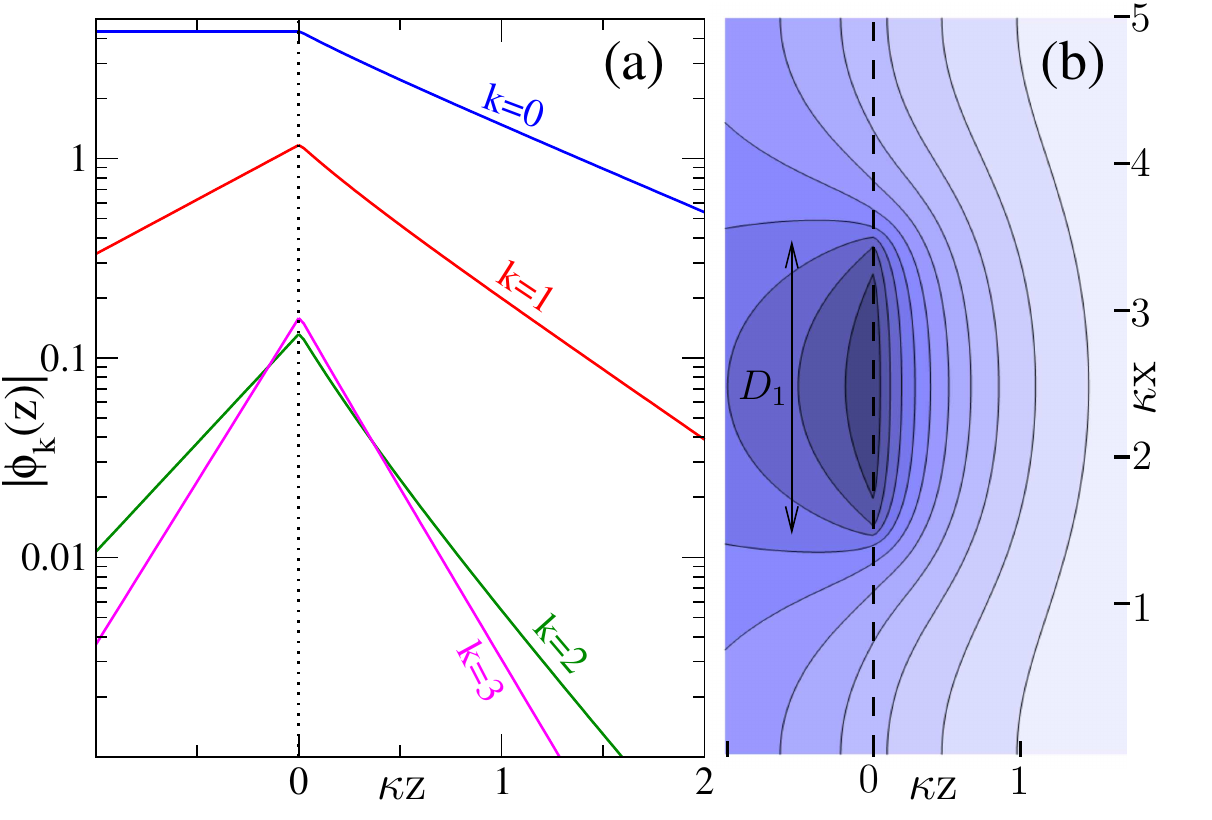}
\caption{Results of the numerical calculation for the electrostatic potential of a striped patchy surface with periodicity $L=5\kappa^{-1}$, stripe width $D_1=2 \kappa^{-1}$ and chargeability $Y=100$, showing the absolute value of the first $4$ mode amplitudes of the electrostatic potential in (a), and a contour plot of the electrostatic potential at both sides of the interface using the calculated mode amplitudes (b). A contour line borders the darkest area at $\Phi(x,z)> 2.75$ and the lightest area at $\Phi(x,z)<0.5$, with steps of $0.25$ in between. The vertical dashed line in both figures denotes the location of the interface between the solid medium (left) and the solvent (right). For these calculations we choose an index-matching plate and liquid, $\epsilon=\epsilon_P$, such that the electric field is able to penetrate into the solid medium.} \label{fig:CONTOURPLOT}
\end{figure}

\Fig~\ref{fig:CONTOURPLOT} shows the electrostatic potential around the plate for $\kappa L=5$, $\kappa D_1 = 2$, $Y_2^\pm = Y_1^-\equiv0$, $Y_1^+=100$ and $\epsilon_P=\epsilon$. This choice of the dielectric constants is such that electric fields do not vanish inside the solid medium. In \fig~\ref{fig:CONTOURPLOT}(a) the logarithmic plot shows nearly exponential decay of the mode amplitudes $\phi_k(z)$ for $k=0,\dots,3$. Deviations from a straight line are due to nonlinear couplings between the modes, and slopes far away from the surface are the mode-dependent screening parameters $\kappa_k$. It can be seen that, for this choice of $\epsilon_P$, the nonzero modes also give rise to an electric field inside the solid medium, and for high $k$ the screening on both sides of the interface is equally efficient, as there are almost no ions involved. \Fig~\ref{fig:CONTOURPLOT}(b) shows a contour plot of the potential $\Phi(x,z)$ around the charged interface. It shows a local increase of the potential close to the stripe, which is caused by a local high charge density. The inhomogeneity in the potential persists for a few screening lengths into the liquid, as the stripe width is larger than the screening length, and thus ionic screening is the dominant type of screening there. In the solid medium ionic screening is not possible, and the inhomogeneity persists over a range of the order of the width of the stripe as was found below \eq~(\ref{modeexp}).\\

\begin{figure}[]
\centering
\includegraphics[width = 8cm]{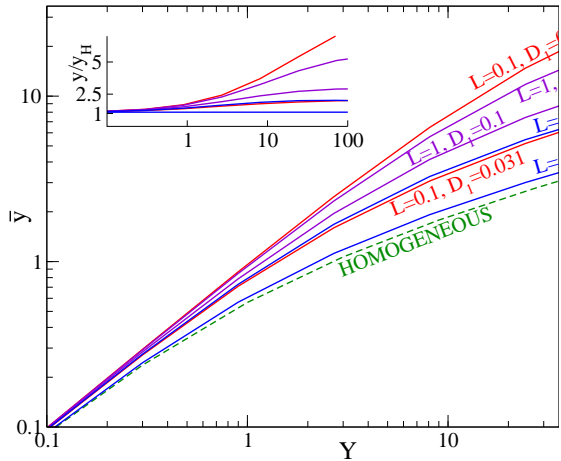}
\caption{Average charge density on a chargeable stripe surrounded by charge-neutral surface for various stripe widths $D_1$ and periodicity $L$, denoted in units of $\kappa^{-1}$, plotted as a function of the stripe chargeability $Y$. The medium has a dielectric constant $\epsilon_P \ll \epsilon$. The dashed line corresponds to the infinite stripe-width limit for which we use the analytical result for homogeneously chargeable plates. Using the same data, the inset shows the charge density relative to this homogeneous limit.} \label{fig:AVCHARGE}
\end{figure}

We now return to the case of chargeable stripes on an otherwise neutral plate, and define the average charge density on a stripe
\begin{equation}
\bar {\y} =  \frac{1}{D_1} \int_{x=0}^{L} \y(x)~\mathrm{d}x. \label{yintegral}
\end{equation}
\Fig~\ref{fig:AVCHARGE} shows $\bar{\y}$ as a function of chargeability $Y$ for multiple choices of the stripe geometry parameters $D_1$ and $D_2$. All curves show, as expected, that the charge increases with the chargeability.
It illustrates, however, that the charging of the stripes strongly depends on their width and the distance in between two successive stripes. Charge densities are increased either by narrower stripes or by larger stripe-stripe distances, the strongest charging occurs for narrow stripes with relatively much space in between. This is intuitively clear since for stripes that are rather separated the charging of the edges is not hindered by the charge of neighboring stripes. Moreover, narrow stripes have relatively more edge surface. For broad stripes this gain at the edges is small compared to the charge at the center, and we see from the line for $\kappa D_1=3.1$ and $\kappa L=10$ that the charge density indeed approaches the homogeneous limit ($D_1=L$), given in \eqs~(\ref{phihom2})-(\ref{yhom}) and limiting behaviour~(\ref{limitingcases}). By contrast, the narrower stripe of width $D_1 = 0.1 \kappa^{-1}$ and the same periodicity $L=10\kappa^{-1}$ has, for $Y\gg1$, a charge density that is a factor $\sim 2$ higher than the homogeneous limit, while the relative increase of $\bar{\y}$ compared to $\y_H$ can be a factor $5$ for extremely narrow patches with $D_1=0.0031\kappa^{-1}$ and $L=0.1\kappa^{-1}$, as can be seen in the inset of \fig~\ref{fig:AVCHARGE}. Furthermore, we note that $\bar{\y}=Y$ in the low-charge limit $Y\ll 1$. This is not surprising either, as in this limit any charge, regardless its position on the surface, is essentially ``alone'' in a vanishingly small potential; the vicinity of a neutral area ``2'' or an essentially neutral stripe ``1'' is indistinguishable in that case.\\

\subsection{Analytical approximation}
\begin{figure}[]
\centering
\includegraphics[width = 8cm]{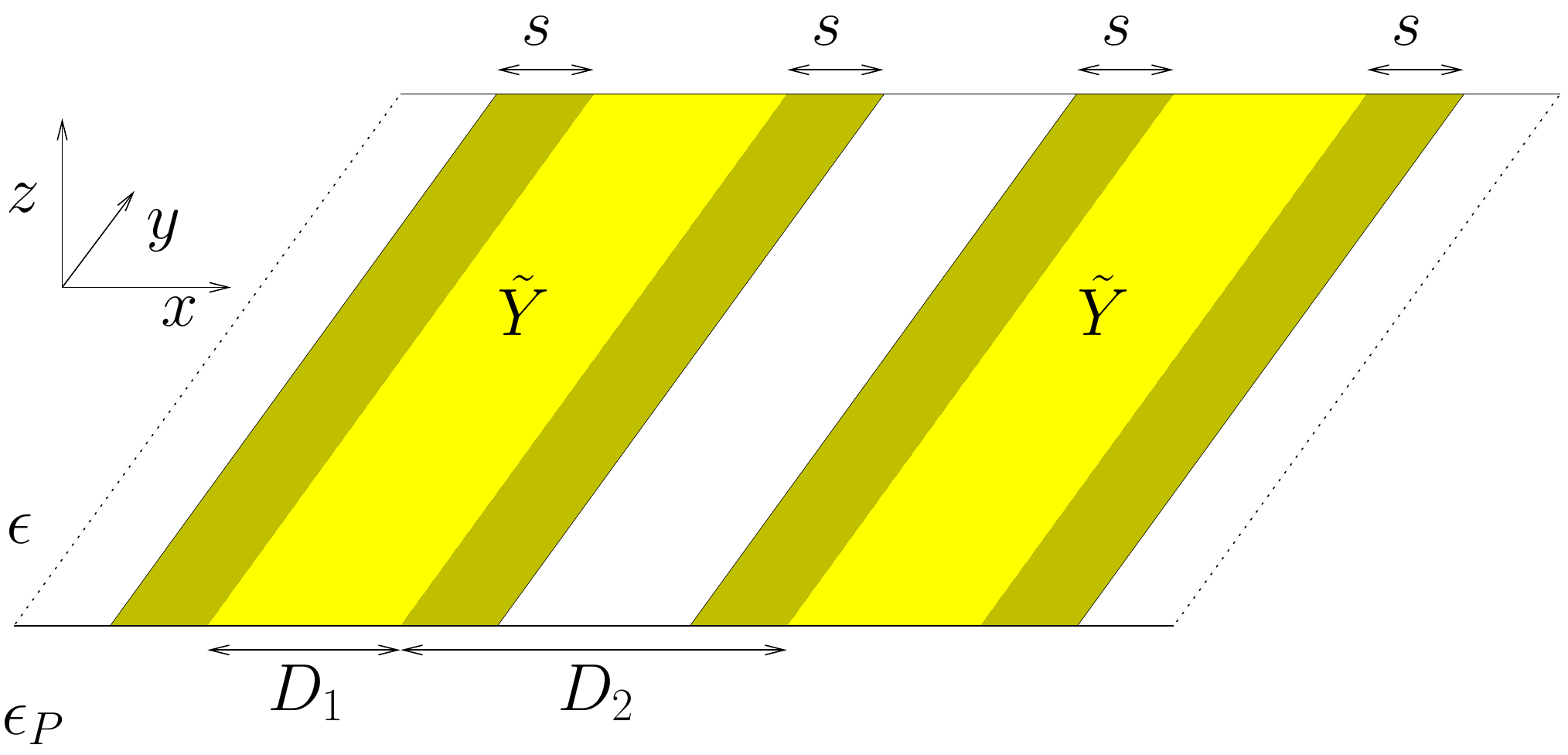}
\caption{Pictorial representation of the method used to estimate the charging of narrow stripes. The darker areas adjoining the shaded stripes resemble strips of charge neutral surface assumed to be involved in the charging as well.\\~\\} \label{fig:METHOD2}
\end{figure}

As a way to better understand and quantify the charging mechanism of the patches, we now propose a method to estimate the stripe charge densities by an analytic procedure and check if the right physics emerges. The key element in the method is the fact that the charging of different surface groups is correlated. This correlation is mediated by the electrostatic potential and logically the (longest) correlation length is of the order of the screening length, as long as we use the assumption that $\epsilon_P \ll \epsilon$. This correlation implies that a small area of charge-neutral plate around every stripe is involved in the charging as well. For the geometry of current interest this means we can think of an extra strip of width $s$ on each side of the charged patch of width $D_1$, as illustrated in \fig~\ref{fig:METHOD2}. Now we presume that the stripe including the extra area charges up like a homogeneously charged plate, such that we can apply the analytic expressions (\ref{phihom2})-(\ref{yhom}) from planar PB theory with an effective (decreased) chargeability $\tilde{Y} = Y \frac{D_1}{D_1 + 2s}$ for the enlarged stripe.\\

The strip width $s$ depends on the local screening length, and therefore we set  $s = \min(\alpha \bar{\kappa}^{-1},D_2/2)$, with $\alpha$ a fit parameter and $\bar{\kappa}^{-1}$ the effective screening length defined below. The minimum condition is used to prevent overlaps, as the size of the additional strip area cannot exceed the size of the uncharged region in between the charged stripes. The effective screening parameter $\bar{\kappa}$ is determined by the average electrostatic potential around the edge. The relation stems from PB-linearization procedures and is given for example in ref. \cite{Alexander, ZoetekouwPRL, ZoetekouwPRE}. Since the local surface potential at the strip must be somewhere in between zero and the value for a homogeneously charged plate with chargeability $Y$, we estimate the effective screening length to be their average, such that $\bar{\kappa}^{-1} = \kappa^{-1} / \sqrt{\cosh{(0+\Phi_{H}(0))/ 2}}$. The procedure is now to use this effective screening length in the calculation of the estimated strip width $s$ to obtain the effective chargeability $\tilde{Y}$. In the homogeneous limit this gives a charge density $\tilde{y}$ for the stripes plus the side strips via \eqs~(\ref{yhom}) and (\ref{phihom}).The average charge density $\bar{\y}$ on the original stripe follows by assigning the charge density to a smaller surface, $\bar{\y} = \tilde{\y} \frac{D_1+2s}{D_1}$.\\

\begin{figure}[]
\centering
\includegraphics[width = 8cm]{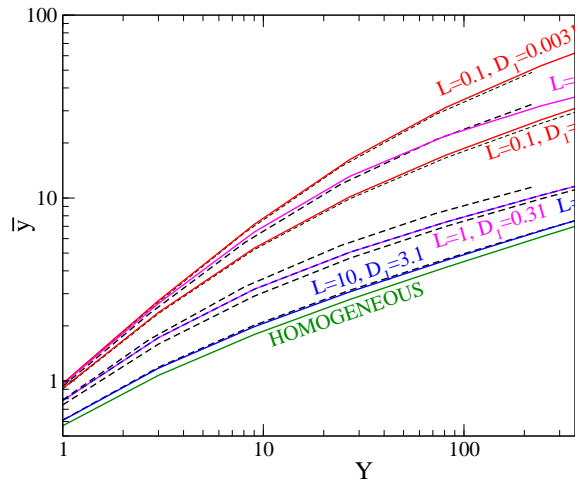}
\caption{Average charge density on a chargeable stripe surrounded by charge-neutral surface, for various stripe widths $D_1$ and separations $D_2$, plotted as a function of the stripe chargeability $Y_1^+$. The medium bares a dielectric constant $\epsilon_P \ll \epsilon$. The solid lines show results from the approximative method as described in the text, whilst the dotted lines show results from the numerical calculations.} \label{fig:AVESTIMATE}
\end{figure}

For the fit parameter $\alpha = 3/8$, \fig~\ref{fig:AVESTIMATE} shows the resulting $\bar{\y}$ as a function of $Y$ for several $\kappa L$ and $\kappa D_1$, together with the numerically determined ``exact'' results based on \eq~(\ref{yintegral}). The agreement between the two is good, within a maximal error of $25$\%, and we find the right trend with $Y$ which suggests that this approximation holds for even higher stripe chargeabilities as well. It is expected that the fits will remain good or even become better for larger stripe distances or stripe widths than investigated here, since for broader stripes the relative amount of edge surface decreases. Moreover, the choice of parameters $\kappa L=10$ in combination with $\kappa D_1 \leq 3.6$ describes a system in which the stripes are separated by multiple screening lengths, which is close to the $\kappa D_2 \rightarrow \infty $ limit where the patches do not mutually interact anymore.

\section{Conclusion}
Although the importance of charge inhomogeneities on the interactions between surfaces was already mentioned in many other studies, the relation between chemical inhomogeneity on the surface and the resulting inhomogeneity in surface charge was, to our knowledge, never studied in detail. We developed a numerical method, based on expansion into Fourier modes, to find the electrostatic potential and charge distributions close to an interface with an inhomogeneous distribution of chargeable chemical surface groups. We focussed on the case where the chemical groups are clustered in stripe-like regions, with different chargeabilities, either of the same or different sign. For the case of striped chargeable patches on an otherwise neutral surface we showed a significant increase of charging of surface groups as the width of the stripes decreases below the Debye length. For very small patches this increase of the surface charge density can easily be an order of magnitude, although patches that have little spacing show significantly less increase of charging because neighboring patches hinder each other's charging. From the numerical results we arrive at the observation that the edges of the patches are able to charge up optimally and will contribute significantly to the total charge of small patches. Our calculated numerical results for the average charge density on a patch are in fine agreement with a very basic analytical model (and one fit parameter) which employs an effective patch size in combination with results for homogeneous plates. Our results are stepping stone towards the study of interactions between heterogeneously charged surfaces, where charged patches may induce charges onto nearby (otherwise) neutral surfaces, thereby generating nontrivial forces and torques. This could be relevant for understanding self-assembly of patchy nanoparticles or proteins. Work along these lines is in progress.\cite{Boon2}\\

This work is financially supported by an ECHO grant of the Nederlandse Organisatie voor Wetenschappelijk Onderzoek (NWO).

%
\end{document}